\documentclass[conference]{IEEEtran}
\IEEEoverridecommandlockouts

\usepackage{cite}
\usepackage{amsmath,amssymb,amsfonts}
\usepackage{algorithmic}
\usepackage{graphicx}
\usepackage{textcomp}
\usepackage{xcolor}
\usepackage{threeparttable}
\usepackage{multirow}
\usepackage{url}

\def\BibTeX{{\rm B\kern-.05em{\sc i\kern-.025em b}\kern-.08em
    T\kern-.1667em\lower.7ex\hbox{E}\kern-.125emX}}
\begin{document}

\title{DrawSpeech: Expressive Speech Synthesis Using  Prosodic Sketches as Control Conditions\\
\thanks{This project is partially supported by the General Research Fund from the Research Grants Council of Hong Kong SAR Government (Project No. 14202623).}
}

\makeatletter
\newcommand{\linebreakand}{%
  \end{@IEEEauthorhalign}
  \hfill\mbox{}\par
  \mbox{}\hfill\begin{@IEEEauthorhalign}
}
\makeatother

\author{\IEEEauthorblockN{Weidong Chen}
\IEEEauthorblockA{
\textit{The Chinese University of Hong Kong}\\
Hong Kong SAR, China \\
wdchen@se.cuhk.edu.hk}
\and
\IEEEauthorblockN{Shan Yang}
\IEEEauthorblockA{\textit{Tencent AI Lab}\\
Shenzhen, China \\
shaanyang@tencent.com}
\and
\IEEEauthorblockN{Guangzhi Li}
\IEEEauthorblockA{\textit{Tencent AI Lab}\\
Shenzhen, China \\
guangzhilei@tencent.com}
\linebreakand
\IEEEauthorblockN{Xixin Wu}
\IEEEauthorblockA{\textit{The Chinese University of Hong Kong}\\
Hong Kong SAR, China \\
wuxx@se.cuhk.edu.hk}
}

\maketitle

\begin{abstract}

Controlling text-to-speech (TTS) systems to synthesize speech with the prosodic characteristics expected by users has attracted much attention. To achieve controllability, current studies focus on two main directions: (1) using reference speech as prosody prompt to guide speech synthesis, and (2) using natural language descriptions to control the generation process. However, finding reference speech that exactly contains the prosody that users want to synthesize takes a lot of effort. Description-based guidance in TTS systems can only determine the overall prosody, which has difficulty in achieving fine-grained prosody control over the synthesized speech. In this paper, we propose DrawSpeech, a sketch-conditioned diffusion model capable of generating speech based on any prosody sketches drawn by users. Specifically, the prosody sketches are fed to DrawSpeech to provide a rough indication of the expected prosody trends. DrawSpeech then recovers the detailed pitch and energy contours based on the coarse sketches and synthesizes the desired speech. Experimental results show that DrawSpeech can generate speech with a wide variety of prosody and can precisely control the fine-grained prosody in a user-friendly manner. Our implementation and audio samples are publicly available\footnote{https://happycolor.github.io/DrawSpeech}.

\end{abstract}


\section{Introduction}

Given a piece of text, we can express it through various prosodic patterns, each conveying distinct meanings \cite{intro}. For example, in the same sentence, speakers may emphasize different words to highlight specific information, leading to entirely different interpretations \cite{emphasis_2}.
Thus, controlling a text-to-speech (TTS) system to synthesize the exact prosody desired by users has attracted considerable attention from researchers \cite{survey, ns1, emo_css, lei_control, Rall-e, Multi_scale_Reference_Encoder}.


Current research on prosody-controllable TTS systems has explored two primary directions. 
One approach involves selecting a reference speech as a prosody prompt and instructing the TTS model to replicate the prosodic pattern on the target text \cite{ref_xutan, Multi_scale_Reference_Encoder, StyleTTS, ns3, ns2, reference_style_control, Text2SE, valle2}.
Typically, StyleTTS \cite{StyleTTS} extracted a style vector from the reference speech and adopted a prosody predictor to predict pitch and energy based on the extracted style vector.
NaturalSpeech 3 \cite{ns3} disentangled prosody attributes from the reference speech, which then served as prosody condition for a prosody diffusion model to generate prosody representations for the input text. 
However, finding a reference speech that precisely matches the desired prosody pattern can be time-consuming \cite{LLM_decision}.
In addition, transferring the prosody from the reference speech results in reduced prosodic variations \cite{seedtts}.

\begin{figure*}[ht]
\centering
\includegraphics[width=0.9\linewidth]{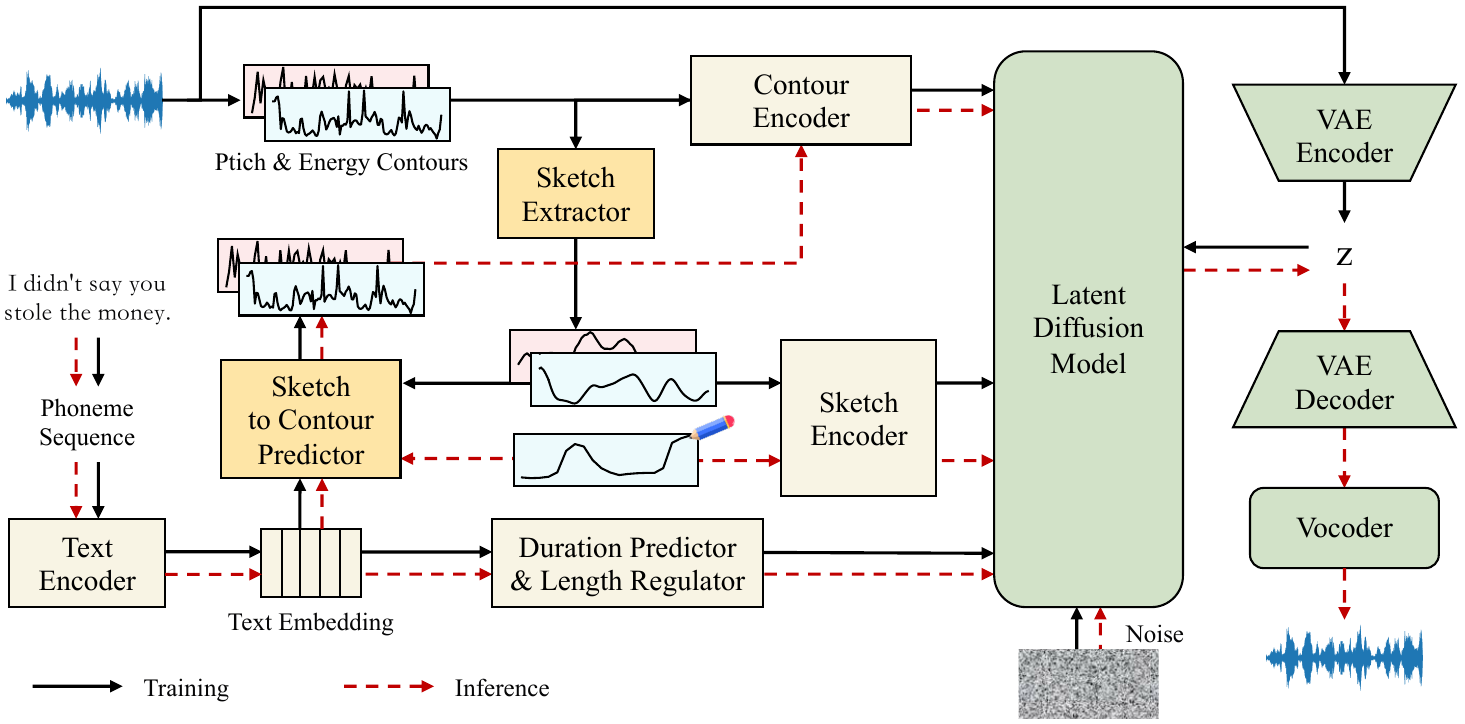}
\caption{Overview structure of the proposed DrawSpeech. Paired speech and text data are used for training. User-supplied text and drawn pitch or energy sketch are used as inputs during inference.}
\label{framework}
\end{figure*}

Another mainstream approach for controlling synthesized prosody
is to use natural language descriptions to specify the desired prosody \cite{InstructTTS, control_nl, PromptStyle, Prompttts, uniaudio, cosyvoice, funaudiollm, text_guided_VC}. For example, PromptStyle \cite{PromptStyle} aligns the description prompt embedding with the style embedding during training, enabling description-based control during inference. 
PromptTTS \cite{Prompttts} leveraged a pre-trained language model BERT \cite{bert} to capture the semantic information in the descriptions, which was subsequently used to guide the speech synthesis process.
However, description-based guidance can only manage utterance-level prosody with descriptions such as ``\textit{his voice has a low pitch}", ``\textit{generate a voice with high volume}", or ``\textit{please use gentle tone}", leaving precise control at a fine-grained level still an unsolved problem.
Moreover, constructing a dataset with descriptions to train such models requires significant effort and a high level of expertise \cite{Prompttts}. 

In this paper, we propose a novel prosody control signal, called prosody sketch, for expressive and controllable TTS. Since pitch and energy are crucial components of prosody \cite{prosody, p_e_matter}, our focus is primarily on controlling these two attributes. To be specific, the sketch serves as a guide, indicating the desired prosody trends to the model, such as a rise and fall in pitch or an increase in energy within a specific region. Users can easily sketch the target prosody for each word, making fine-grained prosody control effortless.
To achieve the new control paradigm, we propose DrawSpeech, which first reconstructs the pitch and energy contours from the abstract sketches, and then employs a diffusion model conditioned on the sketches, contours, and target text to synthesize speech.

The contributions of this work are summarized as follows:
\begin{itemize}
    \item We propose using a novel conditioning signal, called the prosody sketch, which can be easily sketched and enables fine-grained prosody control.
    \item We propose a diffusion-based generative model, named DrawSpeech, which conditions on the sketch and is capable of generating speech with varied prosody.
    \item Experiments show the effectiveness of the sketch-based control paradigm and the DrawSpeech system, especially on its capability to achieve precise prosody control.
\end{itemize}

\section{Methodology}

The architecture of DrawSpeech is shown in Fig.~\ref{framework}, where the sketch extractor and sketch-to-contour predictor are used to achieve the transformation between the sketches and contours. Contour encoder and sketch encoder convert the corresponding inputs into different embedding features. A text encoder equipped with a duration predictor and length regulator handles the input transcript.
A latent diffusion model with a variational autoencoder are employed to generate log-mel spectrogram and a vocoder is used to reconstruct speech waveform. 
Details will be provided in the following subsections.

\subsection{Sketch Extractor}

DrawSpeech aims to allow users to easily control the prosody of the synthesized speech. 
Taking pitch as an example, as shown in Fig.~\ref{c_to_s}(a), the pitch contour is vibrating and changes quickly between the adjacent values. Therefore, controlling prosody based on the contour requires users to provide the detailed fluctuations of pitch or energy, which is inconvenient and sometimes even impractical.
A more feasible solution is to simply ask the user to provide the sketch that reflects the pitch or energy trend in the sentence. 
For example, the pitch sketch shown in Fig.~\ref{c_to_s}(b) tells the model that the sentence should start out at a higher pitch then drop quickly, followed by a gradual rise and arrive at the peaks on certain positions, etc. 
It's much easier for users to draw a pitch sketch as in Fig.~\ref{c_to_s}(b) as a control signal because it relieve users' burden to provide the detailed vibration of the pitch contour as in Fig.~\ref{c_to_s}(a), which will be handled by the model instead.

To obtain the sketch for model training, we first extract the phoneme-level pitch contour $P \in \mathbb{R}^{M}$ and energy contour $E \in \mathbb{R}^{M}$ of the input speech, where $M$ denotes the number of phonemes in the sentence. $P$ and $E$ are then smoothed using Savitzky-Golay filter \cite{filter}, which is useful for smoothing data while preserving the shape of the signal. The resulting pitch and energy sketches are denoted as $P_{ske} \in \mathbb{R}^{M}$ and $E_{ske} \in \mathbb{R}^{M}$, respectively.
As shown in Fig.~\ref{c_to_s}(c), the smoothed sketch effectively captures the variation trends of the contour.

\begin{figure}[t]
\centering
\includegraphics[width=0.9\linewidth]{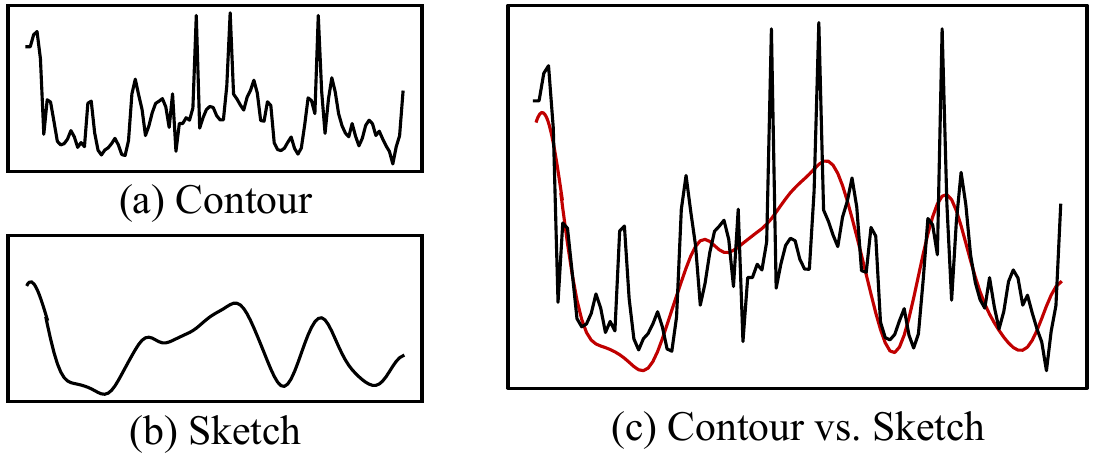}
\caption{Illustrations of (a) pitch contour and (b) pitch sketch. (c) Put the contour and the sketch in the same frame for direct comparison.}
\label{c_to_s}
\end{figure}

\subsection{Sketch-to-Contour Predictor}

While pitch and energy sketches are intuitive for users to comprehend, they omit too much detailed information, making it challenging for speech synthesis systems to reconstruct natural speech.
The sketch-to-contour predictor is designed to leverage text and sketches to restore the detailed contours.

Given the input transcript $y$, we first convert it into the phoneme sequence (with an open-sourced grapheme-to-phoneme toolkit\footnote{https://github.com/Kyubyong/g2p}). Then, a text encoder $f_{text}(\cdot)$ is employed to extract the phoneme embedding $H^y \in \mathbb{R}^{M \times D}$, where $D$ represents the phoneme dimension. The sketch-to-contour predictor can then be written as:
\begin{equation}
    H^y = f_{text}(y)
\end{equation}
\begin{equation}
    P_{pred}, E_{pred} = \mathcal{P}(H^y, P_{ske}, E_{ske})
\end{equation}
where $\mathcal{P}$ represents the sketch-to-contour predictor; $P_{pred} \in \mathbb{R}^{M}$ and $E_{pred} \in \mathbb{R}^{M}$ are the predicted pitch and energy contours, respectively.

\subsection{Latent Diffusion Model}
Firstly, a variational autoencoder (VAE) model \cite{VAE} is employed to compress the mel-spectrogram $x \in \mathbb{R}^{T \times F}$ into a small latent space $z_0 \in \mathbb{R}^{C \times \frac{T}{r} \times \frac{F}{r}}$, where $T$ is the number of frames and $F$ is the frequency dimension, $C$ is the channel dimension and $r$ is the compression rate. 
Simply put, the calculation flow in VAE is as follows:
 \begin{equation}
    z_0 = Enc(x)
\end{equation}
\begin{equation}
    x_{rec} = Dec(z_0)
\end{equation}
where $Enc$ and $Dec$ are the encoder and decoder part of VAE, $x_{rec}$ denotes the reconstructed mel-spectrogram.

Secondly, a contour encoder and a sketch encoder are employed to generate conditions for the latent diffusion model (LDM). 
Specifically, the contour encoder quantizes pitch and energy ($P$ and $E$ during training, and $P_{pred}$ and $E_{pred}$ during inference) into 256 discrete values, subsequently converting them into pitch embedding $H^p \in \mathbb{R}^{M \times F}$ and energy embedding $H^e \in \mathbb{R}^{M \times F}$. The same process takes place in the sketch encoder, except that it utilizes $P_{ske}$ and $E_{ske}$ as inputs and produces pitch sketch embedding $H_{ske}^p \in \mathbb{R}^{M \times F}$ and energy sketch embedding $H_{ske}^e \in \mathbb{R}^{M \times F}$.
To solve the problems of word skipping and repetition, we introduce a duration predictor and a length regulator following \cite{fs}. The ground-truth phoneme durations are used to upsample embeddings during training, while the predicted durations are used during inference. The expanded embeddings are denoted as $\hat{H}^y$, $\hat{H}^p$, $\hat{H}^e$, $\hat{H}_{ske}^p$, $\hat{H}_{ske}^e \in \mathbb{R}^{T \times F}$. Note that a 1D convolutional layer is applied to the phoneme embedding $H^y$ to project the phoneme dimension from $D$ to $F$ prior to the expansion.

Given all the conditions, we employ a LDM \cite{ldm} to generate the latent representation $z_0$ from Gaussian noise $z_N \sim \mathcal N(0,I)$, where $N$ denotes the total sampling steps. For each time step $n \in [1,N]$, the reverse process in LDM is:
\begin{equation}
    \bar{H}^y = \hat{H}^y + \hat{H}^p + \hat{H}^e
\end{equation}
\begin{equation}
    C_n = Concat(z_n, \hat{H}_{ske}^p, \hat{H}_{ske}^e, \bar{H}^y)
\end{equation}
\begin{equation}
    z_{n-1} = z_n - \epsilon_\theta(C_n, t)
\end{equation}
where $Concat(\cdot)$ denotes the concatenation along the channel dimension, $t$ denotes the timestep, $\epsilon_\theta$ denotes the diffusion model with parameters $\theta$. For further details on the forward and reverse processes, as well as the training techniques of the diffusion model, please refer to \cite{ddpm, ldm, audioldm}.

\section{Experiments}

\subsection{Experimental Setup}

\subsubsection{Dataset} 
LJSpeech dataset \cite{ljspeech} is a widely used speech corpus containing 13,100 short audio clips from a single speaker, each paired with a transcription. The total recording duration is roughly 24 hours. We randomly divided the dataset into three subsets: 12,500 samples for training, 300 for validation, and 300 for testing, with a sample rate of 22,050 Hz.

\subsubsection{Implementation Details} 
The text encoder and sketch-to-contour predictor are composed of multiple stacked Transformer \cite{transformer} blocks (6 blocks for the former and 2 blocks for the latter). In both, the linear layer in the feed-forward network is replaced by a 1D convolutional layer. 
The duration predictor and length regulator are identical to those described in \cite{fs}.
The VAE and LDM adopt the structures and configurations used in \cite{audioldm2}.
The contours are normalized using the mean and variance of the entire dataset.
The sketches are normalized to [0, 1].
The dimension of the phoneme embedding $D$ is 256. The window size and hop size for extracting mel-spectrogram are 1024 and 256, respectively. The number of mel frequency bins $F$ is 80. The pre-trained HiFi-GAN\footnote{https://github.com/jik876/hifi-gan} vocoder \cite{hifigan} is utilized to reconstruct speech waveform.

We first train the VAE on LJSpeech for 80k steps with batchsize of 8. 
Subsequently, the VAE and vocoder remain frozen, while the other modules are trained alongside the LDM for 80k steps with batchsize of 32. 
We use the Adam \cite{adam} optimizer to update the VAE and the AdamW \cite{adamw} optimizer to update the LDM along with other modules, and follow the same learning rate schedule in \cite{audioldm2}.
We set either $P_{ske}$ or $E_{ske}$ to an all-zero vector with a probability of 20\% during model training, enabling DrawSpeech to require only a single sketch input from the user during the inference phase. The user-drawn sketch is used as the pitch sketch by default.


\subsubsection{Baselines}

We adopt two TTS models, FastSpeech 2 \cite{fs2} and NaturalSpeech 2 \cite{ns2}, as our baseline models. This selection is based on the fact that both models explicitly incorporate pitch or energy information in speech, allowing us to easily insert the sketch conditions into the original models for comparison.
However, since FastSpeech 2 and NaturalSpeech 2 do not provide official implementations, we instead use the widely adopted public implementation of FastSpeech 2\footnote{https://github.com/ming024/FastSpeech2} and utilize the NaturalSpeech 2 checkpoint provided by Amphion\footnote{https://github.com/open-mmlab/Amphion/tree/main/egs/tts/NaturalSpeech2}.

\subsubsection{Evaluation Metrics}

We use mean opinion score (MOS) to evaluate the sound naturalness, with ratings ranging from 1 to 5 (1-bad, 2-poor, 3-fair, 4-good, 5-excellent).
To evaluate prosody controllability, we introduce sketch correlation (SC), which measures the alignment between the speech and the given sketch. SC is rated on a scale from 1 to 5, with 5 signifying that the sketch accurately reflects the prosody trend of the speech, and 1 indicating no correlation between the given sketch and the synthesized speech. We invite 22 listeners for subjective tests. Note that listeners are instructed to disregard the speech quality when assessing SC. For objective metrics, root mean square error (RMSE) is employed to assess the accuracy of the prosody control.
We select 10 text samples from the test set, with each sample synthesized using two different sketches as prosody condition, resulting in 20 utterances from each system being used for evaluation.

\subsection{Experimental Results}
In this section, we aim to answer three questions: Can other models without special designs utilize sketches as conditions? Will using sketches reduce sound quality? Can using sketches as conditions enable fine-grained and precise prosody control?

\subsubsection{Applying sketches to other models}

To incorporate sketches as conditions into FastSpeech 2,
we directly replace the predicted pitch and energy contours in FastSpeech 2 with the pitch and energy sketches. We also try to multiply the sketches with the predicted contours so that the resulting contours retain the overall trends of the sketches while preserving the details of the original contours. In this case, the control signals are the contours adjusted based on the sketches. For NaturalSpeech 2, since it only predicts the pitch contour, we incorporate only the pitch sketch as condition. The results are shown in Table~\ref{tab_mos_sc}.
From the MOS scores, we can observe that DrawSpeech outperforms its competitors by a substantial margin.
We speculate that the abstract sketches provide insufficient information for FastSpeech 2 and NaturalSpeech 2 to recover natural speech, and the adjusted contours are out-of-domain data for the two baselines.
From the SC scores, we find that DrawSpeech excels at synthesizing speech that closely matches the desired prosody variations, confirming that DrawSpeech can effectively utilize the sketch conditions.
Additionally, when FastSpeech 2 is fine-tuned with specialized modules, the SC score shows a significant improvement compared to the case without fine-tuning, validating the generalization ability of the proposed sketch control method.

\begin{table}[t]
    \caption{MOS and SC comparison with 95\% confidence intervals when adopting sketches (Ske) or contours (Con) as control signals} 
    \label{tab_mos_sc}
    \centering
    \begin{threeparttable}
    \begin{tabular}{l|c|ccc}
    \hline
    Method  &  Control & MOS $\uparrow$ &  SC $\uparrow$ \\ \hline
    \multirow{2}{*}{FastSpeech 2}  & Ske &  2.40 $\pm$ 0.10 & 3.62 $\pm$ 0.09 \\
    & Con\tnote{$\dagger$} & 3.37 $\pm$ 0.08 & 2.82 $\pm$ 0.09 \\ 
    FastSpeech 2 (FT)\tnote{$\ddagger$} & Ske  & 3.55 $\pm$ 0.14 & 3.98 $\pm$ 0.13 \\ \hline
    \multirow{2}{*}{NaturalSpeech 2} & Ske &  2.51 $\pm$ 0.09 & 2.38 $\pm$ 0.10 \\
    & Con\tnote{$\dagger$} & 2.51 $\pm$ 0.09 & 2.35 $\pm$ 0.10 \\ \hline
    DrawSpeech    & Ske  &  \textbf{4.49 $\pm$ 0.06} & \textbf{4.30 $\pm$ 0.07}    \\ \hline
    \end{tabular}
    \begin{tablenotes}
        \footnotesize
        \item[$\dagger$] The contours are adjusted based on the sketches.
        \item[$\ddagger$] We fine-tune FastSpeech 2 with sketch conditions. The Sketch Extractor and Sketch-to-Contour Predictor are integrated into FastSpeech 2 during fine-tuning for a fair comparison.
    \end{tablenotes}
    \end{threeparttable}
\end{table}

\subsubsection{Impact on sound quality}

\begin{table}[t]
    \caption{MOS with 95\% confidence intervals in the ablation studies} 
    \label{tab_mos}
    \centering
    \begin{threeparttable}
    \begin{tabular}{l|cc}
    \hline
    Method & MOS $\uparrow$ & p-value           \\ \hline
    GroundTruth    &  4.46 $\pm$ 0.10  & -\\ \hline
    DrawSpeech with text only   &  3.91 $\pm$ 0.09 & 1.3e-24\\
    DrawSpeech with sketch conditions  &  \textbf{4.49 $\pm$ 0.06} & 0.058 \\ \hline
    \end{tabular}
    \end{threeparttable}
\end{table}

Table~\ref{tab_mos} presents the MOS scores of DrawSpeech with and without the sketch conditions.
The samples generated by DrawSpeech under sketch conditions even slightly surpass the ground truth speech, with a Wilcoxon rank-sum test \cite{ptest} at a p-level of $p > 0.05$.
We conclude that sketch conditions introduce sufficient prosody variations, enhancing the naturalness of the synthetic speech.
The p-value serves as an indicator, demonstrating that DrawSpeech can synthesize speech of high quality or, as defined in \cite{ns1}, that DrawSpeech has achieved human-level quality.

\subsubsection{Precise prosody control}
When a reference speech is provided, we can extract the pitch and energy sketches from reference and use them as conditions to guide the generation process. As shown in Table~\ref{tab_rmse}, DrawSpeech conditioned on the sketches of the reference speech achieves lower pitch and energy RMSE compared to the case without sketch conditions, indicating that DrawSpeech successfully imitate the prosody of the reference speech.
To clearly demonstrate the fine-grained prosody control, we draw four different sketches to emphasize four different words in sentence ``\textit{I didn't say you stole the money}''. The visualization shown in Fig.~\ref{control} suggests that the pitch variations in the synthesized speech closely follow the trends of the corresponding sketches. 
Also, the highest pitch is accurately assigned to the intended emphasis words, indicating that DrawSpeech can achieve prosody control precisely.

\begin{table}[t]
    \caption{RMSE between synthesized speech and reference speech}
    \label{tab_rmse}
    \centering
    \begin{threeparttable}
    \begin{tabular}{l|cc}
    \hline
    \multirow{2}{*}{Method} & \multicolumn{2}{c}{RMSE $\downarrow$} \\ 
    & Pitch (Hz) & Energy (dB)    \\ \hline
    DrawSpeech with text only   &  110.65  & 19.04 \\
    DrawSpeech with sketch conditions  &  \textbf{62.48} & \textbf{9.48} \\ \hline
    \end{tabular}
    \end{threeparttable}
\end{table}

\begin{figure}[t]
\centering
\includegraphics[width=0.9\linewidth]{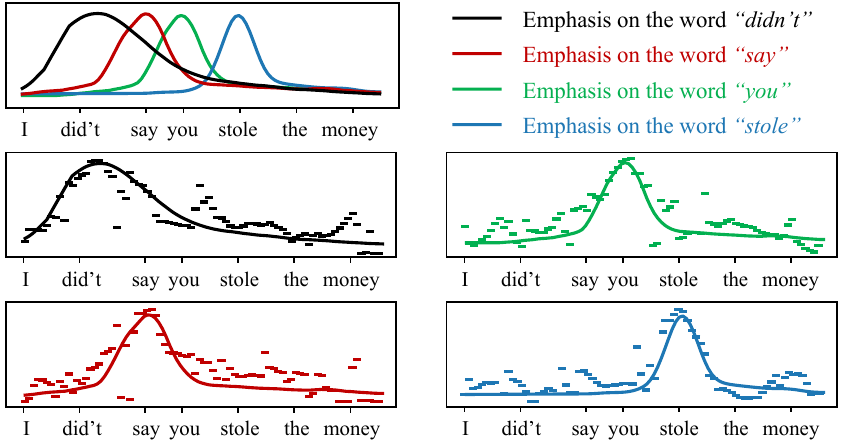}
\caption{Drawing different sketches to achieve precise prosody control. The solid line represents the drawn pitch sketch. The points indicate the pitch of each frame in the synthesized speech.}
\label{control}
\end{figure}

\section{Conclusion}
In this paper, we propose to use the pitch and energy sketches to control the synthesis process. Accordingly, we propose DrawSpeech to generate speech with rich prosody based on the sketches. Experimental results show the effectiveness of DrawSpeech and its capability to achieve precise prosody control. In the future, we plan to incorporate more control signals, such as jitter and shimmer, in a user-friendly manner.

\bibliographystyle{IEEEbib}
\bibliography{refs}

\end{document}